\documentclass[11pt]{article}
\usepackage[a4paper,left=1.5cm,right=1.5cm,top=2cm,bottom=2.5cm]{geometry}
\usepackage{amsmath}
\usepackage{subcaption}
\usepackage{amssymb}
\usepackage{graphicx}
\renewcommand{\vec}{\mathbf}

\begin{document}

\title{Survey on Semi-Explicit Time Integration of\\Eddy Current Problems}
\author{Jennifer Dutin\'{e} 
\and
Markus Clemens 
\and
Sebastian Sch\"{o}ps 
}
\maketitle

\abstract{The spatial discretization of the magnetic vector potential formulation of magnetoquasistatic field problems results in an infinitely stiff differential-algebraic equation system. It is transformed into a finitely stiff ordinary differential equation system by applying a generalized Schur complement. Applying the explicit Euler time integration scheme to this system results in a small maximum stable time step size. Fast computations are required in every time step to yield an acceptable overall simulation time. Several acceleration methods are presented. }

\section{Introduction}
\label{dutine:introduction}
Spatially discretizing the magnetic vector potential formulation of eddy current problems, e.g by the Finite Element Method (FEM), yields a differential-algebraic equation system (DAE) \cite{Schoeps12}. Commonly, only unconditionally stable implicit time integration methods as e.g. the implicit Euler method or the singly diagonal implicit Runge-Kutta schemes can be used for time integration of the infinitely stiff DAE system \cite{HairerWanner}. In every time step at least one large nonlinear algebraic equation system needs to be solved due to the nonlinear BH-characteristic in ferromagnetic materials. The Newton-Raphson method is frequently used for linearization and requires at least one iteration per time step. Here, the Jacobian and the resulting stiffness matrix are updated in each iteration and the resulting linear algebraic equation system needs to be solved efficiently.

Applying explicit time integration schemes avoids the necessity of linearization, because nonlinearities only appear in right-hand side expressions. A first approach to use an explicit time integration method for eddy current problems has been proposed in \cite{Yioultsis01}, where in the conducting regions of the problems the Finite Difference Time Domain (FDTD) method is used. In the nonconducting regions, i.e., in the air, the corresponding parts of the solution are computed using the boundary element method (BEM) \cite{Yioultsis01}.  In a second approach presented in \cite{Ausserhofer09}, the Discontinuous Galerkin FEM and an explicit time integration method are used for computations in the conducting regions. Continuous FEM ansatz functions and an implicit time integration scheme are applied to the nonconducting regions of the problem \cite{Ausserhofer09}.
Both approaches in \cite{Yioultsis01} and \cite{Ausserhofer09} are based on a separate treatment of conducting and nonconducting regions. A different approach presented in \cite{Schoeps12} and \cite{Clemens11} proposes a Schur complement reformulation of the eddy current problem. In\cite{EMF16} the use of a generalized Schur complement is proposed. Here, a pseudo-inverse of the singular curl-curl matrix in nonconducting regions is evaluated using the preconditioned conjugate gradient (PCG) method. This evaluation forms a multiple-right hand side problem  and suitable start vectors for the PCG method are computed using the cascaded Subspace Projection Extrapolation (CSPE) method, which is a modification of the Subspace Projection Extrapolation (SPE) method \cite{Clemens04,EMF16}. Alternatively, the Proper Orthogonal Decomposition (POD) method can be used for computing improved start vectors \cite{CEFC16}. Computations can be accelerated further by using a selective update strategy for updating the reluctivity matrix in conducting regions \cite{IGTE16}. This paper presents a survey on the methods presented in \cite{EMF16,CEFC16,IGTE16}.

\section{Mathematical Formulation}
\label{dutine:mathFormulation}

The partial differential equation
\begin{equation}
\label{dutine:PDE}
\kappa\frac{\partial{\vec{A}(t)}}{\partial t}+\vec{\nabla}\times\Bigl(\nu\bigl(\vec{A}(t)\bigr)\vec{\nabla}\times\vec{A}(t)\Bigr)=\vec{J}_{\mathrm{s}}(t),
\end{equation}describes magnetoquasistatic field problems using the time-dependent magnetic vector potential $\vec{A}(t)$, where $\kappa$ is the electrical conductivity, $\nu$ is the eventually ferromagnetic, i.e., nonlinearly field dependent, reluctivity and $\vec{J}_{\mathrm{s}}(t)$ is the transient source current density.

Discretizing (\ref{dutine:PDE}) in space, e.g. by FEM using edge elements \cite{Monk,Jin}, yields a differential-algebraic equation system (DAE) described by
\begin{equation}
\label{dutine:DAE}
\vec{M}\frac{\mathrm{d}}{\mathrm{d}t}\vec{a}+\vec{K}(\vec{a})\vec{a}=\vec{j}_{\mathrm{s}},
\end{equation}where $\vec{M}$ is the mass-matrix, $\vec{a}$ is the time dependent vector of the magnetic vector potential, $\vec{K}$ is the stiffness-matrix and $\vec{j}_{\mathrm{s}}$ is the vector of the transient source currents.
The degrees of freedom (DoFs) are separated into two vectors $\vec{a}_{\mathrm{c}}$ and $\vec{a}_{\mathrm{n}}$ for conducting and nonconducting material, respectively and (\ref{dutine:DAE}) is re-ordered into

\begin{equation}
\label{eq:dutine:matrixDAE}
\begin{pmatrix}
\vec{M}_{\mathrm{cc}} & 0 \\
0 & 0
\end{pmatrix}
\!
\frac{\mathrm{d}}{\mathrm{d}t}\begin{pmatrix}
\!\!
\vec{a}_{\mathrm{c}}\\
\vec{a}_{\mathrm{n}}
\end{pmatrix}
\!
+
\!
\begin{pmatrix}
\vec{K}_{\mathrm{cc}}\left(\vec{a}_{\mathrm{c}}\right) & \vec{K}_{\mathrm{cn}} \\
\vec{K}^{\mathrm{T}}_{\mathrm{cn}} & \vec{K}_{\mathrm{nn}}
\end{pmatrix}
\!\!
\begin{pmatrix}
\vec{a}_{\mathrm{c}} \\
\vec{a}_{\mathrm{n}}
\end{pmatrix}
\!
=
\!
\begin{pmatrix}
0\\
\vec{j}_{\mathrm{s,n}}
\end{pmatrix}
\!
,
\end{equation} where $\vec{M}_{\mathrm{cc}}$ is the conductivity matrix in conducting regions, $\vec{K}_{\mathrm{cc}}\left(\vec{a}_{\mathrm{c}}\right)$ is the nonlinear part of the reluctivity related stiffness matrix in conducting regions, $\mathbf{K}_{\mathrm{nn}}$ is the part of the curl-curl operator in air, which is singular, and $\mathbf{j}_{\mathrm{s,n}}$ is the source current vector corresponding to excitations in nonconducting regions. $\vec{M}_{\mathrm{cc}}$ is positive definite if using a conventional Galerkin scheme with (possibly high-order) edge elements as test and ansatz functions \cite{Monk, Jin}.
 
The generalized Schur complement expression
\begin{equation}
\label{eq:dutine:Schur}
\vec{K}_{\mathrm{S}}:= \vec{K}_{\mathrm{cn}}\vec{K}^{+}_{\mathrm{nn}}\vec{K}^\top_{\mathrm{cn}},
\end{equation} where $\vec{K}^{+}_{\mathrm{nn}}$ is the matrix representation of a pseudo-inverse of $\mathbf{K}_{\mathrm{nn}}$, is applied to (\ref{eq:dutine:matrixDAE})
and transforms the DAE into an ordinary differential equation (ODE) system 
\begin{eqnarray}
\vec{M}_{\mathrm{cc}}\frac{\mathrm{d}}{\mathrm{d}t}\vec{a}_{\mathrm{c}}+\left(\vec{K}_{\mathrm{cc}}\left(\vec{a}_{\mathrm{c}}\right)-\vec{K}_{\mathrm{S}}\right)\vec{a}_{\mathrm{c}}&=&-\vec{K}_{\mathrm{cn}}\vec{K}^{+}_{\mathrm{nn}}\vec{j}_{\mathrm{s,n}}, \label{eq:dutine:ODE1} \\ 
\vec{a}_{\mathrm{n}} &=& \vec{K}^{+}_{\mathrm{nn}}\vec{j}_{\mathrm{s,n}}-\vec{K}^{+}_{\mathrm{nn}}\vec{K}^{\top}_{\mathrm{cn}}\vec{a}_{\mathrm{c}}, \label{eq:dutine:ODE2}
\end{eqnarray} for the vector $\vec{a}_{\mathrm{c}}$, i.e., the degrees of freedom only situated in conductive material \cite{Clemens11,Schoeps12,EMF16}.
The preconditioned conjugate gradient (PCG) method is used for evaluating a pseudo-inverse of $\vec{K}_{\mathrm{nn}}$ \cite{EMF16}. Alternatively, the singular matrix $\vec{K}_{\mathrm{nn}}$ can be regularized using a grad-div regularization by which $\vec{K}_{\mathrm{nn}}$ is transformed into the discrete Laplacian operator in free space \cite{Clemens11}.
Due to finite stiffness, (\ref{eq:dutine:ODE1}) can be integrated in time using explicit time integration schemes as e.g. the explicit Euler method. Here, in the $m$-th time step the expressions
\begin{eqnarray}
\vec{a}^{m}_{\mathrm{c}} \mathrel{\mathop:} &=& \vec{a}^{m-1}_{\mathrm{c}}+\Delta t\vec{M}^{\mathrm{-1}}_{\mathrm{cc}}\left[-\vec{K}_{\mathrm{cn}}\vec{K}^{+}_{\mathrm{nn}}\vec{j}^{m}_{\mathrm{s,n}}-\left(\vec{K}_{\mathrm{cc}}(\vec{a}^{m-1}_{\mathrm{c}})-\vec{K}_{\mathrm{S}}\right)\vec{a}^{m-1}_{\mathrm{c}}\right], \label{eq:dutine:expTimeStepC} \\
\vec{a}^{m}_{\mathrm{n}} \mathrel{\mathop:} &=& \vec{K}^{+}_{\mathrm{nn}}\vec{j}^{m}_{\mathrm{s,n}}-\vec{K}^{+}_{\mathrm{nn}}\vec{K}^{\top}_{\mathrm{cn}}\vec{a}^{m}_{\mathrm{c}}, \label{eq:dutine:expTimeStepN}
\end{eqnarray}
are evaluated for the degrees of freedom in the conductor domain and in the nonconductive domains consecutively, where $\Delta t $ is the time step size.

Evaluating a pseudo-inverse of $\vec{K}_{\mathrm{nn}}$ and the inverse of $\vec{M}_{\mathrm{cc}}$ in (\ref{eq:dutine:expTimeStepC}) and (\ref{eq:dutine:expTimeStepN}) repeatedly using the PCG method forms multiple right-hand side (mrhs) problems since the matrices involved remain constant. The subspace extrapolation (SPE) method can be used for computing improved start vectors for the PCG method \cite{Clemens04,EMF16}. Solution vectors from $n$ previous time steps are orthonormalized using the modified Gram-Schmidt method and form the linearly independent column vectors of the operator $\vec{V}$.
The projected system
\begin{equation}
\label{eq:dutine:CSPE1}
\vec{V}^{\top}\vec{K}_{\mathrm{nn}}\vec{V}\mathbf{z}=\vec{V}^{\top}\vec{r},
\end{equation} where $\vec{r}$ represents the new right-hand side for the full system, is solved for $\vec{z}\in\mathbb{R}^{n}$ using a direct method.
The linear combination of the column vectors in $\vec{V}$ weighted with the coefficients in $\vec{z}$ yields the improved start vector $\vec{x}_{\mathrm{0,CSPE}}$:
\begin{equation}
\label{eq:dutine:CSPE2}
\vec{x}_{\mathrm{0,CSPE}}:=\vec{V}\vec{z}.
\end{equation} 
Only the last column vector in the operator $\vec{V}$ changes in every time step. Therefore, when computing $\vec{K}_{\mathrm{nn}}\vec{V}$ in (\ref{eq:dutine:CSPE1}), all other matrix-column-vector products evaluated can be reused from previous time steps. This modification of the SPE start vector generation method is referred to as "cascaded SPE" (CSPE).

Alternatively, the proper orthogonal decomposition (POD) method can be used for computing improved start vectors for the PCG method \cite{CEFC16,POD}.
A snapshot matrix $\vec{X}$ is assembled using solutions from previous time steps as column vectors. This matrix is decomposed by the singular value decomposition (SVD) \cite{SVD} into:
\begin{equation}
\label{eq:dutine:SVD}
\vec{X}=\vec{U}\Sigma\vec{V}^{\top},
\end{equation} where $\Sigma$ is a diagonal matrix of the singular values and $\vec{U}$ and $\vec{V}$ are orthogonal matrices.
The first $k$ column vectors of $\vec{U}$ corresponding to the $k$ largest singular values $\sigma_{\mathrm{1}},...,\sigma_{\mathrm{k}}$, for which holds
\begin{eqnarray}
\label{eq:dutine:relationSingValues1}
\sigma_{\mathrm{i}}\geq\sigma_{\mathrm{j}}, \mathrm{for\:} i<j,\\
\label{eq:dutine:relationSingValues2}
\frac{\sigma_{\mathrm{1}}}{\sigma_{\mathrm{k}}}\geq tol_{\mathrm{POD}},
\end{eqnarray}become the column vectors of the reduced matrix $\vec{U}_\mathrm{r}$ with a threshold value $tol_{\mathrm{POD}}$. A threshold value commonly used in practical computations is $tol_{\mathrm{POD}}:=10^{4}$.
The improved start vector $\vec{x}_{\mathrm{0,POD}}$ for the PCG method is computed by
\begin{equation}
\label{eq:dutine:StartVecPOD}
\vec{x}_{\mathrm{0,POD}}:=\vec{U}_{\mathrm{r}}\left[\vec{U}^{\top}_{\mathrm{r}}\vec{K}_{\mathrm{nn}}\vec{U}_{\mathrm{r}}\right]^{\mathrm{-1}}\vec{U}^{\top}_{\mathrm{r}}\vec{K}^{\top}_{\mathrm{cn}}\vec{a}_{\mathrm{c}}.
\end{equation}

The explicit Euler method is only stable for time step sizes $\Delta t$ smaller than a Courant-Friedrich-Levy-type time step size $\Delta t_{\mathrm{CFL}}$ given by \cite{Schoeps12}:

\begin{equation}
\label{eq:dutine:TCFL}
\Delta t_{\mathrm{CFL}}\leq\frac{2}{\lambda_{\mathrm{max}}\left(\vec{M}^{-1}_{\mathrm{cc}}\left(\vec{K}_{\mathrm{cc}}\left(\vec{a}_{\mathrm{c}}\right)-\vec{K}_{\mathrm{S}}\right)\right)},
\end{equation} where the maximum eigenvalue $\lambda_{\mathrm{max}}$ is proportional to
\begin{equation}
\label{eq:dutine:prop}
\lambda_{\mathrm{max}}\left(\vec{M}^{-1}_{\mathrm{cc}}\left(\vec{K}_{\mathrm{cc}}\left(\vec{a}_{\mathrm{c}}\right)-\vec{K}_{\mathrm{S}}\right)\right)\propto\frac{1}{h^{2}\kappa\mu},
\end{equation} assuming non-singularity of $\vec{M}_{\mathrm{cc}}$.  Here, $h$ is the smallest edge length in the mesh, $\kappa$ is the electrical conductivity, and $\mu$ is the permeability. Numerical tests have shown that $1/(h^{2}\kappa\mu)$ unfortunately does not give a sharp estimate of $\lambda_{\mathrm{max}}$, such that the largest eigenvalue has to be computed numerically.

Fine spatial discretization can result in small stable time step sizes, due to (\ref{eq:dutine:TCFL}), that can be in the micro- to nano second range. Considering the dynamics of the usual excitation currents in magnetoquasistatic problems, this corresponds to a strong over-sampling. 
It is assumed that the excitation current does not change significantly between succeeding time steps. Therefore it is expected that the vector $\vec{a}_{\mathrm{c}}$ in (\ref{eq:dutine:expTimeStepC}), (\ref{eq:dutine:expTimeStepN}) also only changes marginally between succeeding time steps. The matrix $\vec{K}_{\mathrm{cc}}\left(\vec{a}_{\mathrm{c}}\right)$ is thus only updated if the change between the vector $\vec{a}^{m}_{\mathrm{c}}$ at the time step $m$ and the vector $\vec{a}^{l}_{\mathrm{c}}$ from the time step $l<m$ at which the matrix $\vec{K}_{\mathrm{cc}}\left(\vec{a}^{l}_{\mathrm{c}}\right)$ has last been updated is larger than a chosen tolerance $tol$, as described by \cite{IGTE16}:
\begin{equation}
\label{eq:dutine:update}
\frac{\lVert\vec{a}^{m}_{\mathrm{c}}-\vec{a}^{l}_{\mathrm{c}}\rVert}{\lVert\vec{a}^{l}_{\mathrm{c}}\rVert}>tol,
\end{equation}where $\|\cdot\|$ denotes the l2-norm. However, depending on the gauging used, a different norm might be more appropriate, e.g. using the magnetic energy norm.

\begin{figure}
	\centering
	\begin{subfigure}[t]{0.4\linewidth}
		\centering
		\includegraphics[height=5cm]{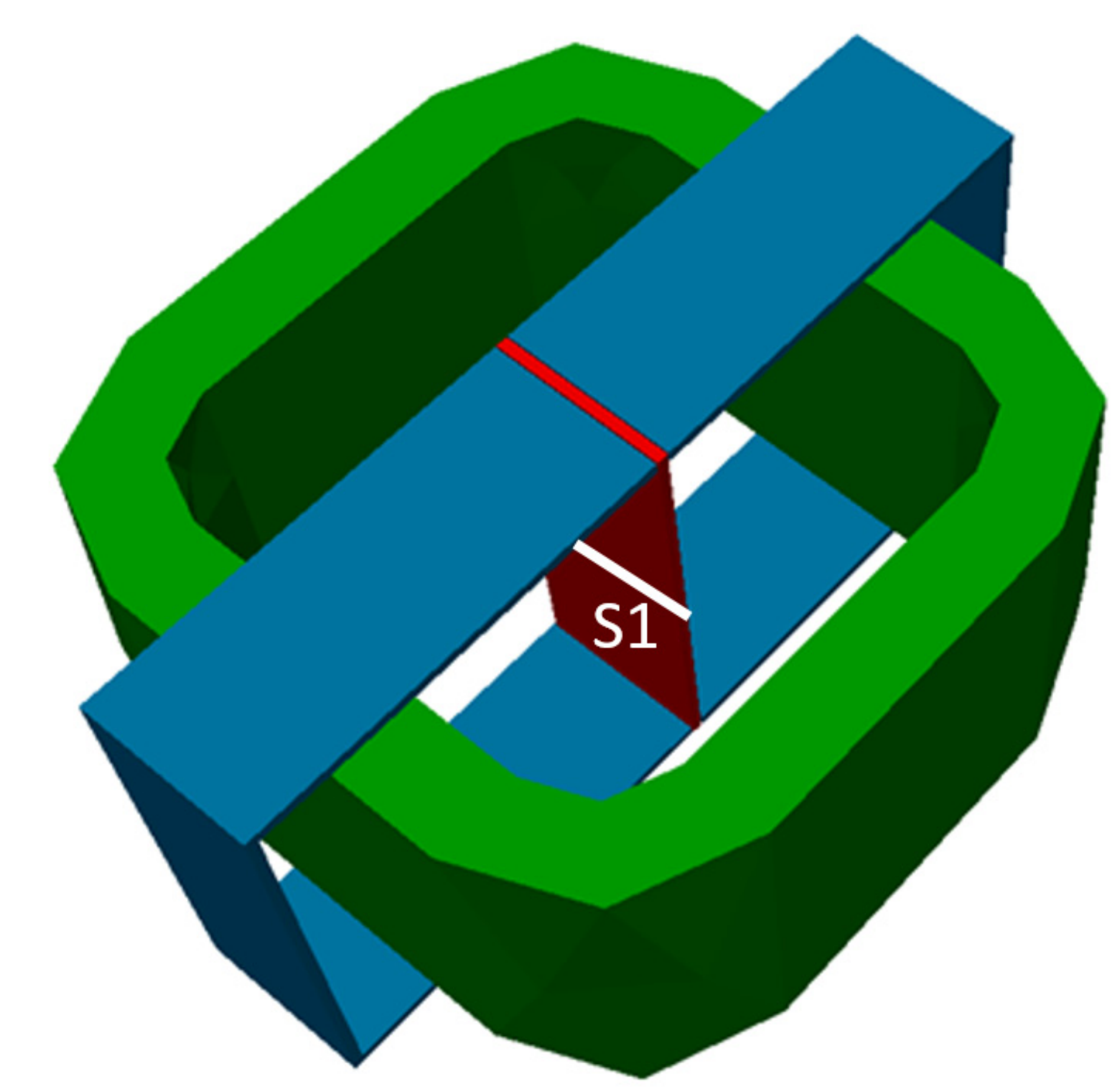}
		\caption{TEAM 10 model geometry and position S1. Steel plates are colored in blue and red, the coil in green. There is a 0.5 mm wide air gap between the blue and red steel plates.}
		\label{fig:dutine:ModelGeometry}
	\end{subfigure}
	\hspace{0.07\linewidth}
	\begin{subfigure}[t]{0.4\linewidth}
		\centering
		\includegraphics[height=5cm]{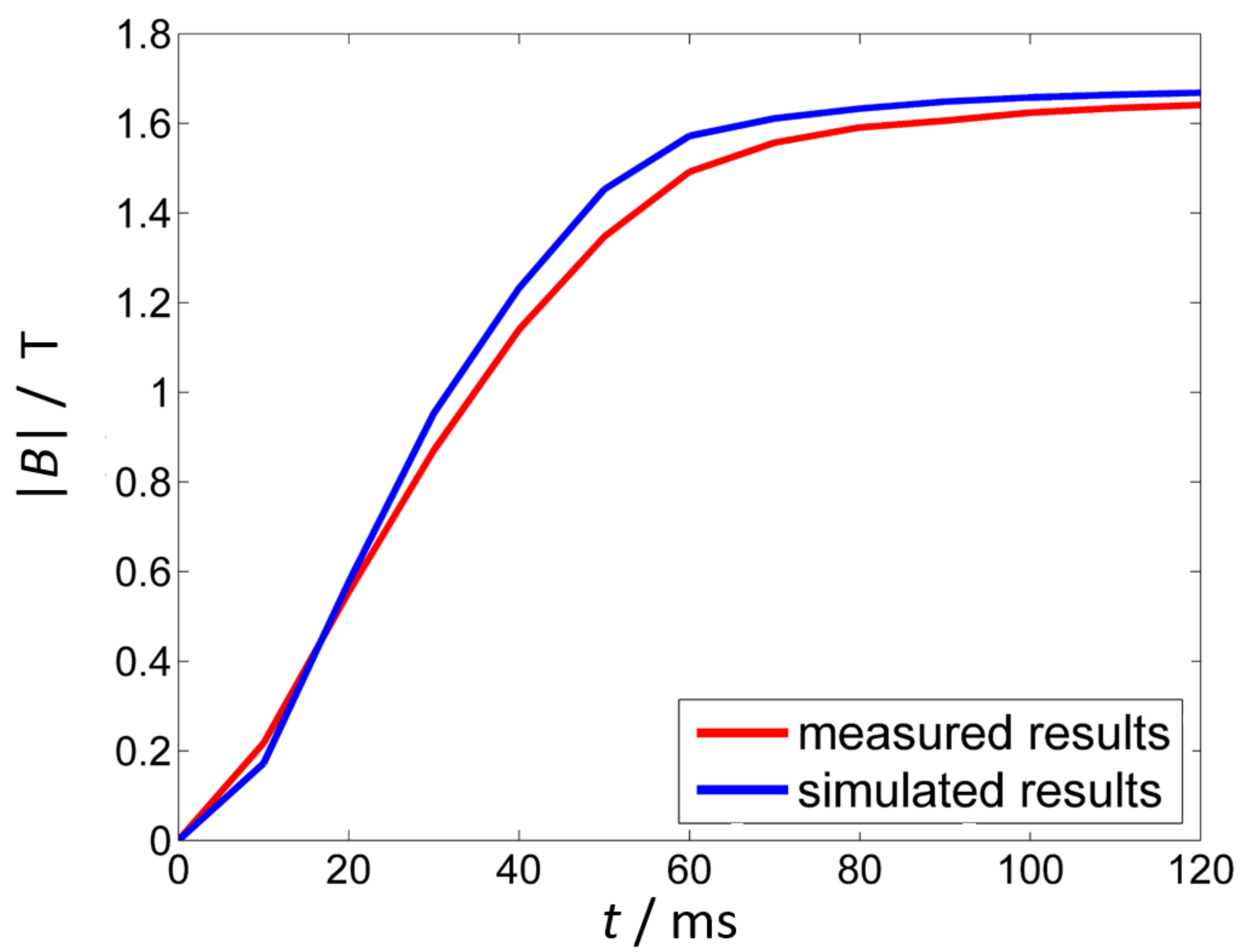}
		\caption{Comparison of simulation results using a mesh of 700,000 DoFs and the measured results published in \cite{Nakata90} at position S1.}
		\label{fig:dutine:ImpMeasured}
	\end{subfigure}
	\caption{TEAM 10 model geometry and comparison of simulation results.}
\end{figure}

\section{Numerical Validation}
The ferromagnetic TEAM 10 benchmark problem is spatially discretized using first order edge element FEM ansatz functions \cite{Nakata90,Kameari90}.
The model geometry is shown in Fig. \ref{fig:dutine:ModelGeometry}. The excitation current is described by a $\left(1-\mathrm{exp}\left(-t / {\tau}\right)\right)$ function. A time interval of $120\,\mathrm{ms}$ duration is calculated. The accuracy of the employed simulation code is proven using an implicit time integration method and a fine mesh discretization of about 700,000 DoFs. The resulting average magnetic flux density is compared with the measurement results published in \cite{Nakata90} in Fig. \ref{fig:dutine:ImpMeasured}. As this simulation takes a simulation time of 5.38 days on a workstation with an Intel Xeon E5 processor, a coarser mesh is applied for the simulations using the explicit Euler method for time integration. The applied coarse spatial discretization yields 29,532 DoFs and results in a maximum stable time step size $\Delta t_{\mathrm{CFL}} = 1.2\,\mu\mathrm{s}$, such that 100,000 explicit Euler time steps are required for this problem.

\begin{figure}[t!]
	\centering
	\begin{subfigure}[t]{0.4\linewidth}
	\centering
	\includegraphics[height=5cm]{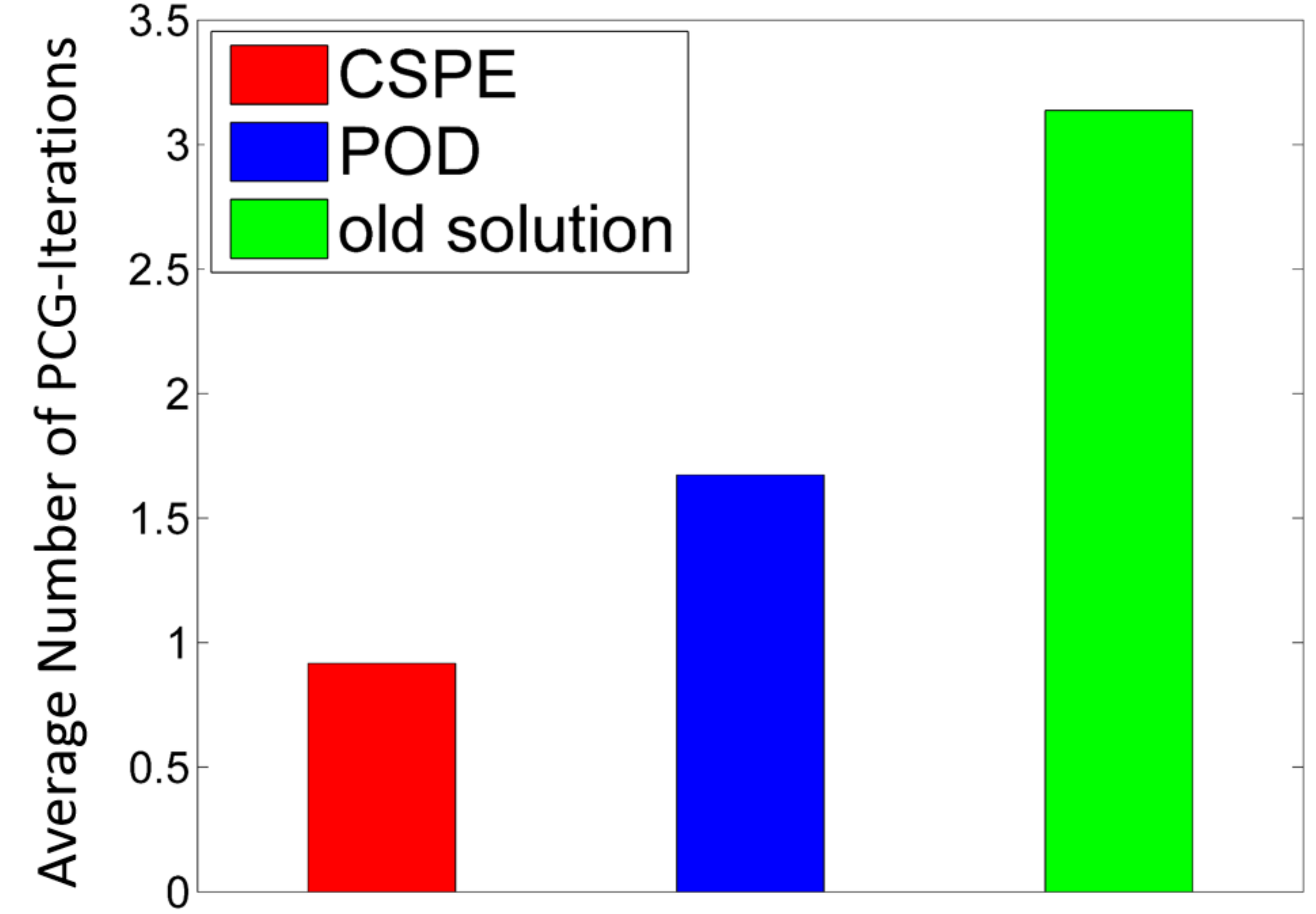}
	\caption{Averagely required number of PCG iterations for evaluating the pseudo-inverse of $\vec{K}_{\mathrm{nn}}$ using either CSPE, POD, or the solution from the previous time step as start vector for the PCG method.}
	\label{fig:dutine:CGIterations}
	\end{subfigure}
	\hspace{0.07\linewidth}
	\begin{subfigure}[t]{0.4\linewidth}
		\centering
		\includegraphics[height=5cm]{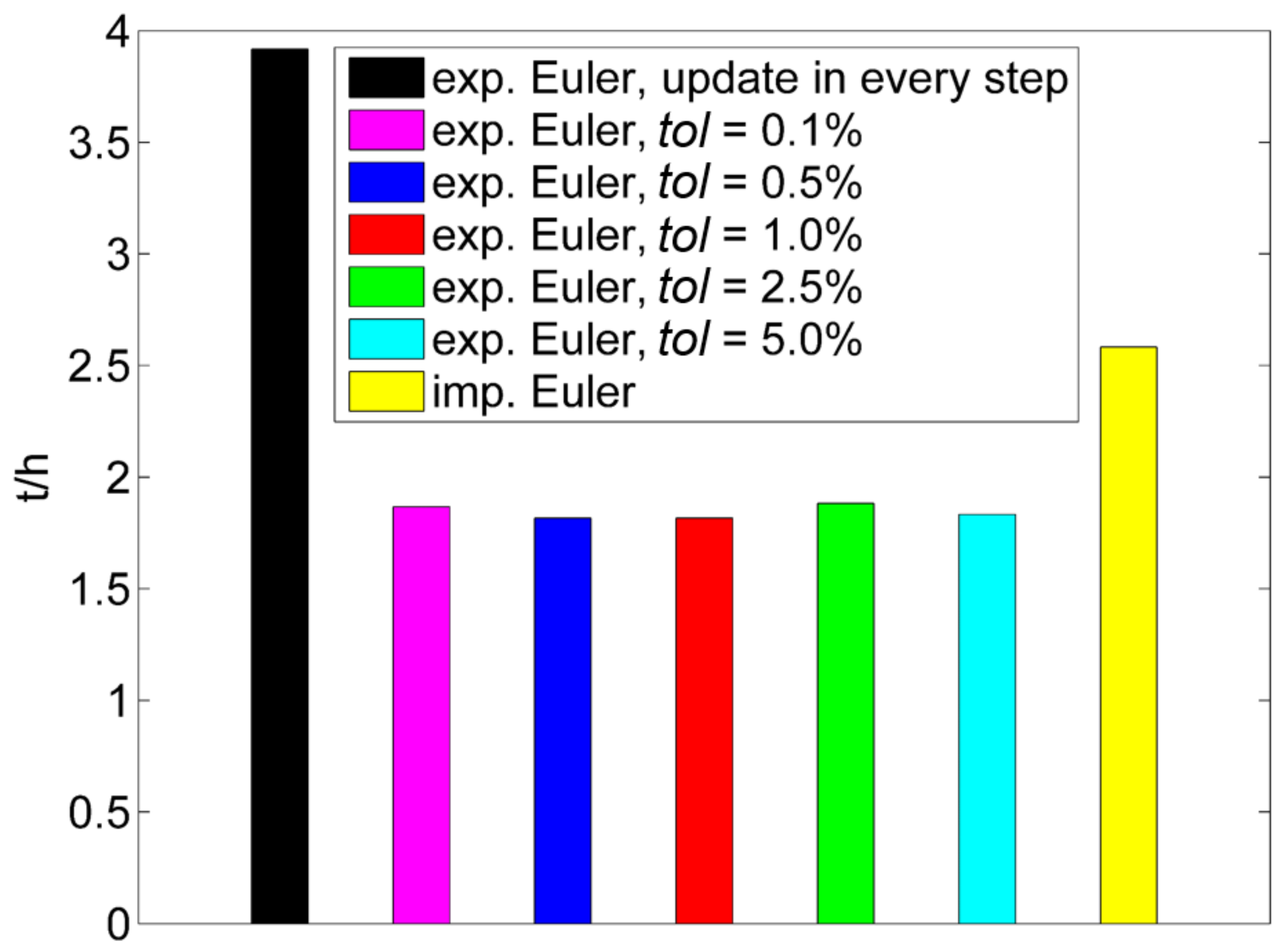}
		\caption{Comparison of simulation times for a simulation using implicit Euler method, the explicit Euler  method with updates of $\vec{K}_{\mathrm{cc}}\left(\vec{a}_{c}\right)$ in every time step and the explicit Euler method using the selective update strategy and different tolerances $tol$.}
		\label{fig:dutine:simTimes}
	\end{subfigure}
	\\[0.5cm]
	\begin{subfigure}[t]{0.4\linewidth}
		\centering
		\includegraphics[height=5cm]{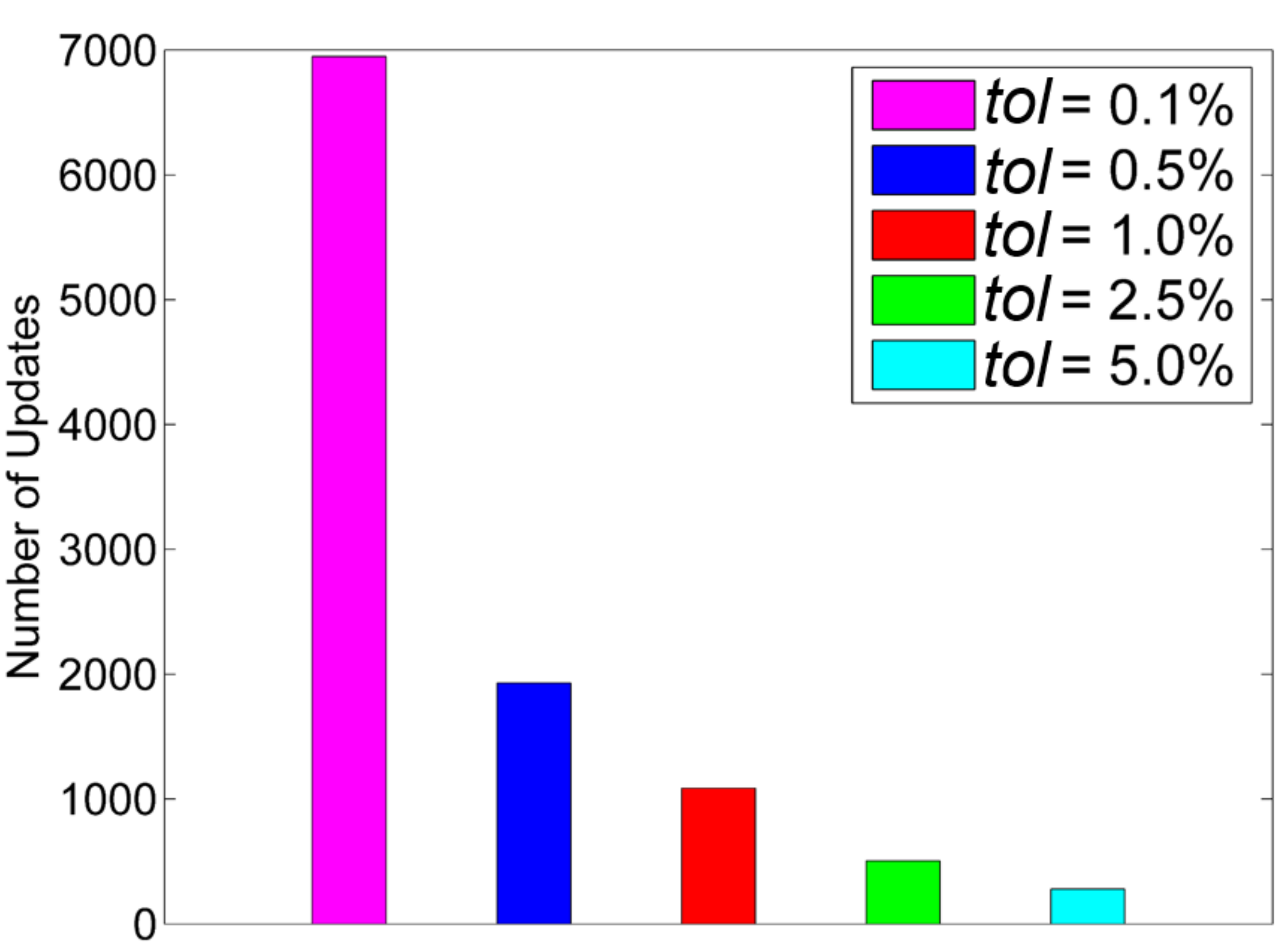}
		\caption{Number of updates of $\vec{K}_{\mathrm{cc}}\left(\vec{a}_{c}\right)$ for different tolerances $tol$ in (\ref{eq:dutine:update}).}
		\label{fig:dutine:numUpdates}
	\end{subfigure}
	\hspace{0.07\linewidth}
	\begin{subfigure}[t]{0.4\linewidth}
	\centering
	\includegraphics[height=5cm]{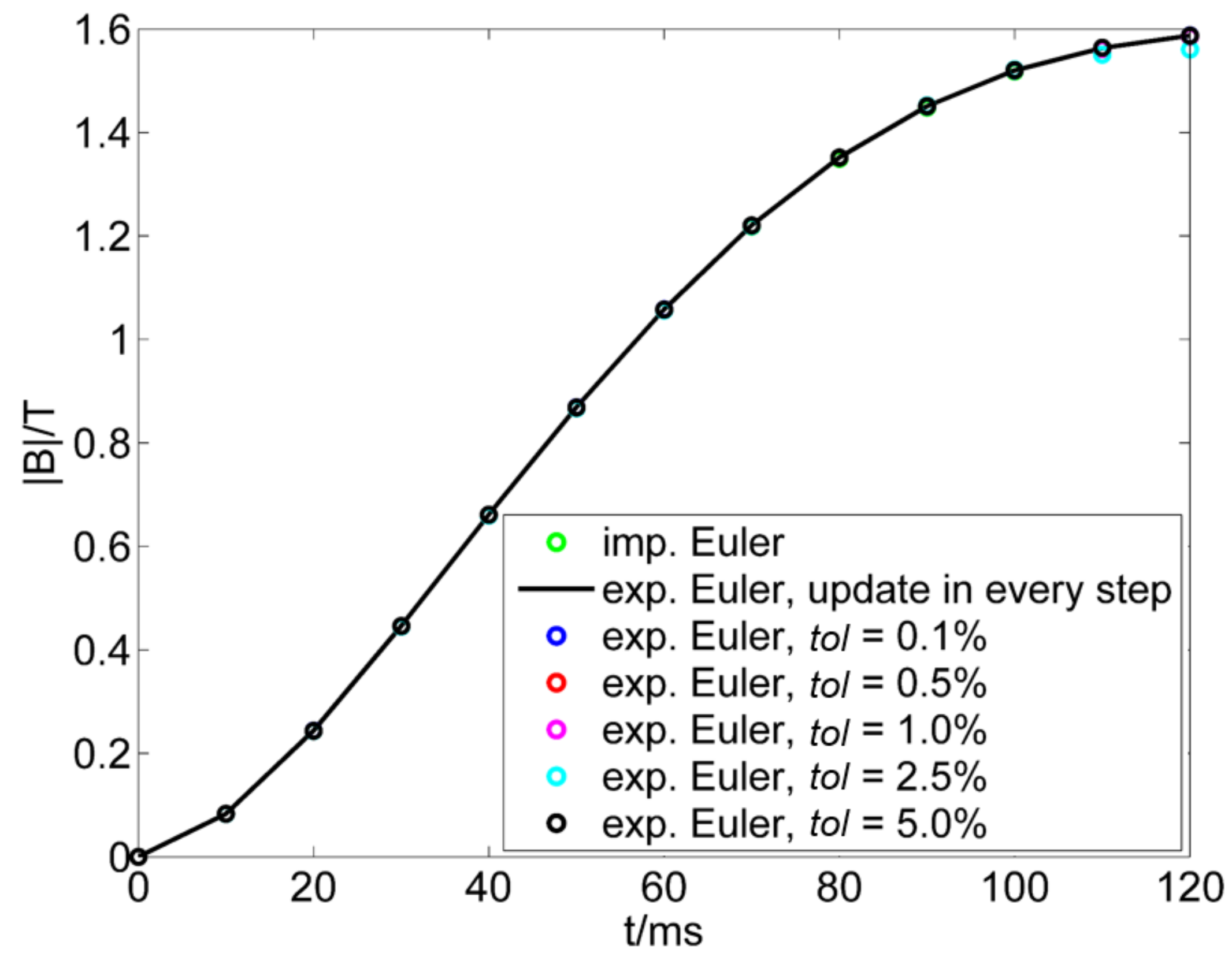}
	\caption{Average magnetic flux density at position~S1.}
	\label{fig:dutine:Bupdates}
	\end{subfigure}
	\caption{Numerical results for implicit and various variants of the explicit Euler method.}
\end{figure}

Computing improved start vectors for the PCG method using either CSPE or POD reduces the average number of required PCG iterations compared to using the solution from the previous time step as start vector. An algebraic multigrid method is used as preconditioner. The results for the evaluation of the pseudo-inverse of $\vec{K}_{\mathrm{nn}}$ using a PCG tolerance of $10^{-6}$ are shown in Fig. \ref{fig:dutine:CGIterations}. Using the selective update strategy for updating the matrix $\vec{K}_{\mathrm{cc}}\left(\vec{a}_{c}\right)$ does not significantly decrease accuracy, as is shown in Fig. \ref{fig:dutine:Bupdates}. The number of required updates and the simulation time are significantly reduced, as is depicted in Fig. \ref{fig:dutine:numUpdates} and Fig. \ref{fig:dutine:simTimes}. If $\vec{K}_{\mathrm{cc}}\left(\vec{a}_{c}\right)$ is updated in every time step 100,000 updates are performed during the entire simulation.
A workstation with an Intel Xeon E5 processor and an NVIDIA TESLA K80 GPU are used for these simulations. The matrix $\vec{M}_{\mathrm{cc}}$ is inverted directly using GPU acceleration. This is only possible, as the matrix $\vec{M}_{\mathrm{cc}}$ is only of dimension 5955x5955 in this test problem. For more refined discretizations the PCG method should be used for inverting the matrix $\vec{M}_{\mathrm{cc}}$. 

\section{Conclusion}
The application of a generalized Schur complement to the spatially discretized magnetic vector potential formulation of magnetoquasistatic field problems transformed a DAE of infinite stiffness into a finitely stiff system of ODEs. This ODE system is integrated with the explicit Euler method. For the evaluation of a pseudo-inverse the PCG method was adopted. Improved start vectors were  computed with the CSPE and the POD method, reducing the number of required PCG iterations in simulations of the ferromagnetic TEAM 10 benchmark problem. A selective update strategy for the reluctivity matrix taking into account the specific problem dynamics reduced the number of required updates and the simulation time.
So far, the small stable time step size of the explicit Euler method results in high computational effort which can be overcome using massive GPU-parallelization to reduce the required computational time per time step significantly.

\section*{Acknowledgement}
This work was supported by the German Research Foundation DFG (grant numbers CL143/11-1 and SCHO1562/1-1). The third author is supported by the Excellence Initiative of the German Federal and State Governments and The Graduate School of Computational Engineering at TU Darmstadt.




\begin{thebibliography}{99} 
\bibitem{Schoeps12} Sch\"{o}ps, S., Bartel, A., Clemens, M.:
Higher order half-explicit time integration of eddy current problems using domain substructuring.
IEEE Trans. Mag., \textbf{48(2)}, 623--626 (2012)

\bibitem{HairerWanner} Hairer, E., Wanner, G.:
Solving Ordinary Differential Equations II: Stiff and Differential-Algebraic Problems (2nd rev. edn.).
Springer, Berlin  (1996)

\bibitem{Yioultsis01} Yioultsis, T.V., Charitou, K.S., Antonopoulos, C.S., Tsiboukis, T.D.:
A finite difference time domain scheme for transient eddy current problems.
IEEE Trans. Mag., \textbf{37(5)}, 3145--3149 (2001)

\bibitem{Ausserhofer09} Au\ss{}erhofer, S., B\'{i}ro, O., Preis, K.:
Discontinuous galerkin finite elements in time domain eddy-current problems.
IEEE Trans. Mag., \textbf{45(3)}, 1300--1303 (2009)

\bibitem{Clemens11} Clemens, M., Sch\"{o}ps, S., De Gersem, H., Bartel, A.:
Decomposition and regularization of nonlinear anisotropic curl-curl daes.
Compel, \textbf{30(6)}, 1701--1714 (2011)


\bibitem{EMF16} Dutin\'{e}, J., Clemens, M., Sch\"{o}ps, S., Wimmer, G.:
Explicit time integration of transient eddy current problems.
presented at the 10th International Symposium on Electric and Magnetic Fields (EMF2016),
full paper submitted.

\bibitem{Clemens04} Clemens, M., Wilke, M., Schuhmann, R., Weiland, T.:
Subspace projection extrapolation scheme for transient field simulations.
IEEE Trans. Mag., \textbf{40(2)}, 934--937 (2004)


\bibitem{CEFC16} Dutin\'{e}, J., Clemens, M., Sch\"{o}ps, S.:
Multiple right-hand side techniques in semi-explicit time integration methods for transient eddy current problems,
accepted for presentation at the 17th Biennial IEEE Conference on Electromagnetic Field Computation (CEFC 2016).


\bibitem{IGTE16} Dutin\'{e}, J., Clemens, M., Sch\"{o}ps, S.:
Explicit Time Integration of Eddy Current Problems Using a Selective Matrix Update Strategy.
accepted for presentation at the 17th International IGTE Symposium on Numerical Field Calculation in Electrical Engineering (IGTE 2016).

\bibitem{Monk} Monk, P.: Finite element methods for Maxwell’s equations.
Oxford University Press, Oxford (2003)

\bibitem{Jin} Jin, J.-M.:
The finite element method in electromagnetics (3rd Edition), 
Wiley-IEEE Press, Hoboken (2014)


\bibitem{POD}
Chatterjee, A.:
An introduction to the proper orthogonal decomposition.
Current Sci., \textbf{78(7)}, 808--817 (2000)

\bibitem{SVD}
Trefethen, L.N., Bau, D.:
Numerical Linear Algebra.
Society for Industrial and Applied Mathematics, Philadelphia (1997) 

\bibitem{Nakata90} Nakata, T., Fujiwara, K.:
Results for benchmark problem 10 (steel plates around a coil).
Compel, \textbf{9(3)}, 181--192 (1990)


\bibitem{Kameari90} Kameari, A.:
Calculation of transient 3d eddy current using edge-elements.
IEEE Trans. Mag., \textbf{26(2)}, 466--469 (1990)



















%
%
%






\end{thebibliography}
\end{document}